\documentclass[12pt]{article}
\pdfoutput =1
\usepackage{float} 
\usepackage{amsmath}
\usepackage{slashed}
\usepackage{amsfonts}   
\usepackage{amssymb}

\begin{document}
\date{}
\title{{\bf{\Large Nonrelativistic spinning strings}}}
\author{
 {\bf {\normalsize Dibakar Roychowdhury}$
$\thanks{E-mail:  dibakarphys@gmail.com, dibakar.roychowdhury@ph.iitr.ac.in}}\\
 {\normalsize  Department of Physics, Indian Institute of Technology Roorkee,}\\
  {\normalsize Roorkee 247667, Uttarakhand, India}
\\[0.3cm]
}

\maketitle
\begin{abstract}
We construct nonrelativistic spinning string solutions corresponding to $ SU(1,2|3) $ Spin-Matrix theory (SMT) limit of strings in $ AdS_5 \times S^5 $. Considering various nonrelativistic spinning string configurations both in $ AdS_5 $ as well as $ S^5 $ we obtain corresponding dispersion relations in the strong coupling regime of SMT where the strong coupling ($ \sim \sqrt{\mathfrak{g}} $) corrections near the BPS bound have been estimated in the slow spinning limit of strings in $ AdS_5 $. We generalize our results explicitly by constructing three spin folded string configurations that has two of its spins along $ AdS_5 $ and one along $ S^5 $. Our analysis reveals that the correction to the spectrum depends non trivially on the length of the NR string in $ AdS_5 $. The rest of the paper essentially unfolds the underlying connection between $ SU(1,2|3) $ Spin-Matrix theory (SMT) limit of strings in $ AdS_5 \times S^5 $ and the nonrelativistic Neumann-Rosochatius like integrable models in 1D. Taking two specific examples of NR spinning strings in $ R \times S^3 $ as well as in certain sub-sector of $ AdS_5 $ we show that similar reduction is indeed possible where one can estimate the spectrum of the theory using 1D model. 
\end{abstract}
\section{Overview and Motivation}
\subsection{Spin-Matrix theory and NR strings}
During last one decade, a series of work \cite{Harmark:2006di}-\cite{Harmark:2019zkn} have been put forward which essentially argues about a more tractable limit of the AdS$ _5 $/CFT$ _4 $ correspondence \cite{Maldacena:1997re}-\cite{Witten:1998qj} in a regime,
\begin{eqnarray}
\mathcal{H}=\lim_{\lambda \rightarrow 0}\frac{\Delta -J}{\lambda}=fixed~;~N=fixed
\label{e1}
\end{eqnarray}   
where both the descriptions are under control and hence the corresponding spectrum could be subjected to a precise test. Here, $ \Delta $ is the energy of a state in $ \mathcal{N}=4$ SYM on $ \mathbb{R} \times S^3 $ and $ J $ is a linear sum over Cartan charges. The above limit (\ref{e1}) essentially takes us towards a (nonrelativistic) corner of $ \mathcal{N}=4 $ SYM known as the Spin-Matrix theory (SMT) \cite{Harmark:2014mpa} where states in the Hilbert space carry indices both in the spin as well as adjoint representation. 

On the gauge theory side of the duality, the above limit (\ref{e1}) corresponds to operators with classical/tree level dimension $ \Delta_0 = J $ where all the other operators with $ \Delta_0 > J $ are essentially decoupled from the rest of the spectrum. As for example, one can think of the $ SU(2) $ sector of $ \mathcal{N}=4 $ SYM where the one loop correction to the scaling dimension ($ \Delta $) could be formally expressed as,
\begin{eqnarray}
\Delta = J + \lambda \Delta_{1} + \mathcal{O}(\lambda^{3/2}).
\label{e3}
\end{eqnarray}

Combining (\ref{e1}) and (\ref{e3}) the corresponding SMT Hamiltonian turns out to be,
\begin{eqnarray}
\mathcal{H}_{SMT}=J+\mathfrak{g}\lim_{\lambda \rightarrow 0}\frac{\Delta -J}{\lambda} =J+ \mathfrak{g} \Delta_{1}
\label{e4}
\end{eqnarray}
where, $ \mathfrak{g} $ is the coupling constant appearing in the SMT limit \cite{Harmark:2014mpa}. Therefore, in the decoupling/SMT limit, the gauge theory side of the duality is fully under control provided one knows the one loop contribution to the anomalous dimension \cite{Harmark:2008gm}. 

In the planar ($ N \rightarrow \infty $) limit, the weakly coupled ($ \mathfrak{g}\ll 1 $) version of the theory recast itself as an integrable spin chain with nearest neighbour interactions \cite{Harmark:2014mpa}. The low energy spectrum of the theory contains excitations like single spin magnons and so on. However, a detailed understanding of similar phenomena from the perspective of strong coupling ($ \mathfrak{g}\gg 1 $) physics is still lacking in the literature. The understanding of strong coupling phenomena requires a dual nonrelativistic (NR) stringy counterpart  \cite{Gomis:2000bd}-\cite{Bergshoeff:2019pij}  living in $ AdS_5 \times S^5 $ which for our case would correspond to constructing a null reduced sigma model action over torsional Newton-Cartan (TNC) geometry \cite{Harmark:2017rpg}-\cite{Roychowdhury:2020kma} and thereby taking $ 1/c $ limit of the world-sheet degrees of freedom. 

On the string theory side of the correspondence, one typically starts with the null reduced form of the metric \cite{Harmark:2017rpg}-\cite{Harmark:2018cdl} (in the presence of a null isometry direction $ \mathfrak{u} $),
\begin{eqnarray}
ds^2 = 2 \tau (d \mathfrak{u}-\mathfrak{m})+\mathfrak{h}_{\mu \nu}dX^{\mu}dX^{\nu}
\end{eqnarray}
corresponding to $ AdS_5 \times S^5 $ and consider the SMT limit (\ref{e4}) as \cite{Harmark:2018cdl},
\begin{eqnarray}
\mathcal{H}_{string}=\lim_{g_s \rightarrow 0}\frac{E -Q}{g_s}=fixed~;~N=fixed~;~Q=S+J
\label{ee6}
\end{eqnarray}   
where, $ g_{s}=\frac{\lambda}{4 \pi N} $ is the string coupling constant. As a matter of fact, in order to construct a finite sigma model action in the above limit (\ref{ee6}) one needs to simultaneously rescale the world-sheet vielbein ($ \tau_{\alpha} $) as well as the string tension ($ T $) with some appropriate factor/power of $ c=\frac{1}{\sqrt{4 \pi g_s N}} $ which goes to infinity in the limit (\ref{ee6}). As a result of this rescaling, the original sigma model action essentially boils down into a NR sigma model action which is conjectured to be dual to the SMT physics as mentioned above \cite{Harmark:2017rpg}-\cite{Harmark:2018cdl}.
\subsection{Summary of results}
One of the primary ambitions of the present paper is to construct various NR (semi-classical) spinning string configurations in $ AdS_5 \times S^5 $ and explore the corresponding dispersion relations in the (large $ \mathfrak{g} \gg 1$) $ SU(1,2|3) $ SMT limit\footnote{ $ SU(1,2|3) $ SMT limit of $ \mathcal{N}=4 $ SYM is of particular interest as this contains $ 1/16 $ BPS supersymmetric states. In the limit of strong ($ \mathfrak{g}\rightarrow \infty $) coupling and large temperatures, these supersymmetric states are supposed be describing supersymmetric black holes on the stringy side of the correspondence. } of $ \mathcal{N}=4 $ SYM. In particular, we construct the NR analogue of GKP like solutions \cite{Gubser:2002tv} by computing the energy as well as spin associated with various folded (spinning) string configurations in $ AdS_5 \times S^5 $ \cite{Frolov:2003xy}-\cite{Ryang:2004tq}. Unlike its relativistic cousins \cite{Gubser:2002tv}, here the results are obtained in the \emph{slow spinning} ($ S \ll \sqrt{\mathfrak{g}} $) limit of NR strings in $ AdS_5 $ which has a dispersion relation of the form,
\begin{eqnarray}
E_{NR}\sim S+\sqrt{\mathfrak{g}}~\sum_{n\geq 2} \mathfrak{a}_n \left( \frac{S}{\sqrt{\mathfrak{g}}}\right)^n 
\label{e6}
\end{eqnarray}
where, the coefficients ($\mathfrak{a}_n$) of the expansion (\ref{e6}) are non trivial functions of string length.\\\\
$ \bullet $ Considering spinning strings in $ S^5 $, on the other hand, we notice that the dispersion relation takes the following form,
\begin{eqnarray}
E_{NR} \sim J +\gamma ~\sqrt{\mathfrak{g}}
\end{eqnarray}
where, $ \gamma $ stands for the leading order correction to the spectrum in the semiclassical limit. Our analysis reveals that the correction becomes large as we move form single spin to multi-spin string configurations.\\\\
$ \bullet $ We generalize above observations by considering \emph{folded} three spin solutions that has two of its spins along $ AdS_5 $ and the remaining one along $ S^5 $. In the slow frequency limit, this configuration yields the dispersion relation that is of the form,
\begin{eqnarray}
E_{NR}\sim Q +\frac{\sqrt{\mathfrak{g}}\rho^2_m}{8 \pi}(\tilde{\Delta}_1 (\rho_m)+\omega  \Sigma_{(\omega)}(\rho_m)+\nu  \Sigma_{(\nu)} (\rho_m)+\cdots) +\mathcal{O}(\sqrt{\mathfrak{g}}/\rho_m)
\label{E8}
\end{eqnarray}
where, $ Q=S+J $ is the total spin of the configuration such that, $ \frac{Q}{\sqrt{\mathfrak{g}}}<1 $. Here, the entities within the parenthesis on the R.H.S. of (\ref{E8}) are some complicated functions of the string length whose detailed expressions have been provided in the following Section 2.\\ \\
$ \bullet $ The rest of the paper essentially unfolds the relationship between $ SU(1,2|3) $ SMT limit of spinning strings in $ AdS_5 \times S^5 $ and the Neumann-Rosochatius integrable systems \cite{Babelon:1992rb} in 1D. The 1D model thus obtained are defined as the nonrelativistic analogue of Neumann-Rosochatius like models those constructed previously in the context of type IIB strings propagating in $ AdS_5 \times S^5 $ \cite{Arutyunov:2003uj}-\cite{Arutyunov:2003za}. Eventually, we compute the spectrum of spinning strings using this reduced model. We consider two specific examples. 

The first example takes into account spinning strings in $ R \times S^3 $ and its 1D reduction to nonrelativistic \emph{extended} Neumann-Rosochatius like systems. The spectrum that one computes from this reduced model typically takes the form,
\begin{eqnarray}
E_{NR}\sim \left( \frac{\mathfrak{m}^2_1}{2}+\frac{\mathfrak{m}^2_2}{3}\right) J +\frac{\sqrt{\mathfrak{g}}}{2}\left(\mathfrak{m}^2_1 +\frac{2}{3}\mathfrak{m}^2_2 - 2\mathfrak{m}^2_2 (\tilde{J}+6 \tilde{J}^2) \right) 
\end{eqnarray}
where, $ \tilde{J}=\frac{J}{\sqrt{\mathfrak{g}}} $ is the effective R- charge in the strong coupling regime of SMT. On the other hand, $ \mathfrak{m}_{1,2} $ are the windning numbers of the string along two of the azimuthal directions ($ \phi_{1,2} $) of $ S^5 $. 

A similar analysis for \emph{folded} spinning strings in $ AdS_5 $ reveals,
\begin{eqnarray}
E_{NR} \sim (2 k^2 - \omega)S_{\varphi} +\sqrt{\mathfrak{g}}\left( \frac{3}{\pi^2}(k^2 - \omega)^2\right)^{1/3} \left( \frac{S_{\varphi}}{\sqrt{\mathfrak{g}}}\right)^{1/3} +\cdots
\label{E10}
\end{eqnarray} 
where, $ k $ is the winding number of the string along one of the azimuthal directions ($ \varphi $) of $ S^3 \subset AdS_5 $ and $ \omega $ is the corresponding spinning frequency of the soliton. While obtaining the above relation (\ref{E10}), one essentially considers the so called \emph{short} string limit where the string soliton is considered to be sitting near the north pole ($ \psi \sim 0 $) of the three sphere ($ S^3 $) and is located near the centre of $ AdS_5 $ as well.

Finally, we conclude in Section 5 where we briefly outline possible interpretation for these nonrelativisitc (spinning) string states in terms of dual SMT degrees of freedom. In particular, we discuss various decoupling limits associated to $ \mathcal{N}=4 $ SYM those may be interpreted as the degrees of freedom of a nonrelativistic string in the limit of strong coupling. This finally provides a platform for several non trivial checks in nonrelativistic holographic correspondence using quantum mechanical degrees of freedom. 
\section{$ SU(1,2|3) $ SMT limit and NR strings}
We start with the null reduced form of the NR sigma model action over $ AdS_5 \times S^5 $ in the $ SU(1,2|3) $ SMT limit. The NR Nambu-Goto (NG) action could be formally expressed as,
\begin{eqnarray}
\mathcal{S}_{NG}=\frac{\sqrt{\mathfrak{g}}}{4\pi}\int d^2 \sigma \mathcal{L}_{NG}~;~\sigma^{\alpha}=\lbrace \sigma^0 , \sigma^1 \rbrace
\end{eqnarray}
where the corresponding Lagrangian density is given by \cite{Harmark:2017rpg}-\cite{Harmark:2018cdl},
\begin{eqnarray}
\mathcal{L}_{NG}=\epsilon^{\alpha \beta}\mathfrak{m}_{\alpha}\partial_{\beta}\eta +\frac{\epsilon^{\alpha \alpha'}\epsilon^{\beta \beta'}\tau_{\alpha'}\tau_{\beta'}}{2 \epsilon^{\gamma \gamma'}\tau_{\gamma}\partial_{\gamma'}\eta}\mathfrak{h}_{\alpha \beta}
\label{e8}
\end{eqnarray}
where $ \eta $ is the compact dual dimension along which the string has a nonzero winding mode. Here $\lbrace \tau_{\alpha}, \mathfrak{m}_{\alpha}\rbrace  $ are the world-sheet one forms and $ \mathfrak{h}_{\alpha \beta} $ is the world-sheet two form whose detailed expressions are given below \cite{Harmark:2018cdl},
\begin{eqnarray}
\tau_{\alpha}=\cosh^2\rho \partial_{\alpha}t
\end{eqnarray}
\begin{eqnarray}
\mathfrak{m}_{\alpha}=-\tanh^2\rho \left( \partial_{\alpha}\chi +\frac{1}{2}\cos\psi \partial_{\alpha}\varphi\right)~~~~~~~~~~~~~~~ \nonumber\\
+\cosh^{-2}\rho\left(\frac{1}{2}\cos\theta_2 \sin^2\theta_1 \partial_{\alpha}\phi_1 -\left( \frac{1}{3}-\frac{1}{2}\sin^2\theta_1\right)\partial_{\alpha}\phi_2 \right) 
\end{eqnarray}
and,
\begin{eqnarray}
\mathfrak{h}_{\alpha \beta}= \tanh^2\rho  \partial_{\alpha}\chi \partial_{\beta}\chi +\partial_{\alpha}\rho \partial_{\beta}\rho +\frac{1}{4}\sinh^2 \rho(\partial_{\alpha}\psi \partial_{\beta}\psi +\sin^2\psi \partial_{\alpha}\varphi \partial_{\beta}\varphi)~~~~~~~~~~~\nonumber\\
+ \tanh^2\rho \partial_{\alpha}\chi \left(\frac{1}{2}\cos\psi \partial_{\beta}\varphi +\frac{1}{2}\cos\theta_2 \sin^2\theta_1 \partial_{\beta}\phi_1 -\left( \frac{1}{3}-\frac{1}{2}\sin^2\theta_1\right)\partial_{\beta}\phi_2 \right)
+(\alpha\leftrightarrow \beta)\nonumber\\
+\frac{\tanh^2\rho}{4}\cos^2\psi \partial_{\alpha}\varphi \partial_{\beta}\varphi +\partial_{\alpha}\theta_1 \partial_{\beta}\theta_1 +\frac{1}{4}\sin^2\theta_1 (\partial_{\alpha}\theta_2 \partial_{\beta}\theta_2 +\sin^2\theta_2 \partial_{\alpha}\phi_1 \partial_{\beta}\phi_1)\nonumber\\
+\frac{1}{4}\sin^2 \theta_1 \cos^2\theta_1 (\partial_{\alpha}\phi_2 \partial_{\beta}\phi_2 + \cos\theta_2 (\partial_{\alpha}\phi_1 \partial_{\beta}\phi_2 +\partial_{\beta}\phi_1 \partial_{\alpha}\phi_2 )+ \cos^2\theta_2 \partial_{\alpha}\phi_1 \partial_{\beta}\phi_1)\nonumber\\
+\tanh^2\rho \left( \frac{\cos\psi}{2}\left(\frac{1}{2}\cos\theta_2 \sin^2\theta_1 \partial_{\alpha}\phi_1 -\left( \frac{1}{3}-\frac{1}{2}\sin^2\theta_1\right)\partial_{\alpha}\phi_2 \right) \partial_{\beta}\varphi  +(\alpha\leftrightarrow \beta)\right) \nonumber\\
+\tanh^2\rho \left(\frac{1}{2}\cos\theta_2 \sin^2\theta_1 \partial_{\alpha}\phi_1 -\left( \frac{1}{3}-\frac{1}{2}\sin^2\theta_1\right)\partial_{\alpha}\phi_2 \right) \nonumber\\
\times \left(\frac{1}{2}\cos\theta_2 \sin^2\theta_1 \partial_{\beta}\phi_1 -\left( \frac{1}{3}-\frac{1}{2}\sin^2\theta_1\right)\partial_{\beta}\phi_2 \right).
\end{eqnarray}

Notice that, the angles $ \psi $ and $ \varphi $ belong to $ S^3 $ which is a subset of the full $ AdS_5 $ geometry. The information regarding the remaining coordinate of the three sphere is encoded in the angular direction $ \chi $ which is periodic with a period of $ 2\pi $. On the other hand, the remaining angular variables $ \lbrace \theta_1, \theta_2, \phi_1, \phi_2\rbrace $ belong to the five sphere ($ S^5 $) part of the original 10D target space geometry.
\subsection{Spinning strings in $ AdS_5 $}
We consider closed NR spinning strings in $ AdS_5 $ and construct solutions with single spin in $ S^3 $. To work with such stringy configurations, we choose the following ansatz
\begin{eqnarray}
t = \sigma^{0} ~;~ \rho = \rho (\sigma^{1})~;~\varphi = \omega  \sigma^{0} ~;~ \eta =  \sigma^{1} ~;~\psi = \psi ( \sigma^{1})
\end{eqnarray}
and switch off all the remaining coordinates.

The corresponding NG Lagrangian density (\ref{e8}) takes the following form,
\begin{eqnarray}
\mathcal{L}_{NG}=-\frac{ \omega}{2}\tanh^2 \rho \cos\psi +\frac{1}{2}\cosh^2\rho\left( \rho'^2 +\frac{\psi'^2}{4}\sinh^2\rho\right)
\end{eqnarray}
where prime corresponds to derivative w.r.t. $ \sigma^1 $.

The resulting equations of motion may be obtained as,
\begin{eqnarray}
\label{e14}
\rho'' \cosh^2\rho +\rho'^2 \cosh\rho \sinh\rho + \omega \tanh\rho \cosh^{-2}\rho \cos\psi - \frac{\psi'^2}{16}\sinh4\rho &=&0\\
\psi'' \cosh^2\rho \sinh^2\rho +\frac{\psi' \rho'}{2}\sinh4\rho -2 \omega \tanh^2\rho \sin\psi &=&0.
\label{e15}
\end{eqnarray}
\subsubsection{Short strings}
To proceed further, we choose to work with the  \emph{short} string \cite{Gubser:2002tv} limit where we consider that the (folded) string is not stretched enough and is located near the centre  ($ \rho \sim 0 $) of $ AdS_5 $. In other words, this is the limit that essentially describes NR spinning strings in \emph{flat} space where we ignore the curvature effects of $ AdS_5 $. We also consider that the center of the string soliton is located at the north pole ($ \psi =0 $) of $ S^3 $. Combining all these pieces together we finally obtain,  
\begin{eqnarray}
\rho'' + \omega \rho  & \approx & 0
\end{eqnarray}
where we retain ourselves only upto quadratic order in the bosonic field fluctuations $ \rho(\sigma^1) $. The solution corresponding to the radial fluctuation may be obtained as,
\begin{eqnarray}
\rho (\sigma^1)\sim \rho_m \sin \left(\sigma^{1}  \sqrt{\omega }\right)
\end{eqnarray}
where $ \rho_m (\ll1) $ is the maximum value of the radial coordinate such that the string has four segments in it, each of which is ranging between 0 to $ \rho_m $.

A straightforward computation further reveals the energy,
\begin{eqnarray}
E_{NR} =\frac{\sqrt{\mathfrak{g}}}{4\pi }\int_{0}^{2\pi} d\sigma^1 \frac{\delta \mathcal{L}_{NG}}{\delta \dot{t}} \approx \frac{\sqrt{\mathfrak{g}}}{2\pi}\int_{0}^{\rho_m} d\rho \rho' = \frac{\sqrt{\mathfrak{g}}}{8}\sqrt{\omega}\rho^2_m
\label{e19}
\end{eqnarray}
as well as the spin angular momentum,
\begin{eqnarray}
S_{\varphi}=\frac{\sqrt{\mathfrak{g}}}{4\pi }\int_{0}^{2\pi} d\sigma^1 \frac{\delta \mathcal{L}_{NG}}{\delta \dot{\varphi}}\approx \frac{\sqrt{\mathfrak{g}}}{2\pi}\int_{0}^{\rho_m} d\rho \frac{\rho^2}{\rho'}=\frac{\sqrt{\mathfrak{g}}}{8}\frac{\rho^2_m}{\sqrt{\omega}}.
\label{e20}
\end{eqnarray}

Combing (\ref{e19}) and (\ref{e20}) we finally obtain,
\begin{eqnarray}
E_{NR} \sim \omega S_{\varphi}.
\end{eqnarray}
\subsubsection{Extended strings}
In order to extract dispersion relation corresponding to NR extended (spinning) strings we again start with the following equation that describes strings sitting at the north pole of $ S^3 $,
\begin{eqnarray}
\rho'' \cosh^2\rho +\rho'^2 \cosh\rho \sinh\rho + \omega \tanh\rho \cosh^{-2}\rho =0.
\end{eqnarray}

In order to simplify the problem, we look into the specific sector of the parameter space namely we try to construct solution in the slow spinning limit of the string. In the slow velocity ($ \omega \ll 1/\rho_m $) limit, the corresponding solution may be expressed as,
\begin{eqnarray}
\rho (\sigma^1)=\rho_m \sinh ^{-1}\sigma^1 \left( 1+ \frac{\omega(2\sigma^1 +\tan ^{-1}\sigma^1) }{2 \sqrt{(\sigma ^1)^2+1} \sinh ^{-1}\sigma^1}+\cdots\right).
\end{eqnarray}

The energy of the configuration may be noted down as,
\begin{eqnarray}
E_{NR}= \frac{\sqrt{\mathfrak{g}}}{8\pi}\int_{0}^{2\pi}d\sigma^1 \rho'^2 \cosh^2\rho \simeq \frac{\sqrt{\mathfrak{g}}}{8\pi}(\Delta_1 (\rho_m)+\omega \rho_m^3 \Delta_{2}(\rho_m)+\cdots) 
\label{e23}
\end{eqnarray}
which after a straightforward computation yields\footnote{Here, $ \, _2\tilde{F}_1\left(a,b ;c ;d\right) $ is the \emph{regularised} hypergeometric function \cite{abramh}.},
\begin{eqnarray}
\Delta_1 (\rho_m) = \, _2\tilde{F}_1\left(1,\frac{1}{2}-\rho_m ;\frac{3}{2}-\rho_m ;-4 \pi  \left(\sqrt{1+4 \pi ^2}+2 \pi \right)-1\right) \nonumber\\
\times e^{(1-2 \rho_m) \sinh ^{-1}(2 \pi )} \Gamma \left(\frac{1}{2}-\rho_m \right)\frac{\rho^2_m}{4}\nonumber\\
+ _2\tilde{F}_1\left(1,\rho_m +\frac{1}{2};\rho_m +\frac{3}{2};-4 \pi  \left(\sqrt{1+4 \pi ^2}+2 \pi \right)-1\right)\nonumber\\
\times e^{(2 \rho_m +1) \sinh ^{-1}(2 \pi )} \Gamma \left(\rho_m +\frac{1}{2}\right)\frac{\rho^2_m}{4}\nonumber\\
+\frac{\rho^2_m}{4}\left( 2 \tan ^{-1}(2 \pi ) - \pi  \sec (\pi  \rho_m)\right)
\end{eqnarray}
and,
\begin{eqnarray}
\Delta_2 (\rho_m) =\int_0^{2 \pi }d\sigma^1 \frac{ \left(2 \sigma^1 +\tan ^{-1}\sigma^1 \right) }{2\left((\sigma^1) ^2+1\right)^{3/2}}\sinh \left(2\rho_m \sinh ^{-1}\sigma^1 \right). 
\end{eqnarray}

The spin angular momentum, on the other hand, turns out to be,
\begin{eqnarray}
S_{\varphi}=\frac{\sqrt{\mathfrak{g}}}{8\pi}\int_{0}^{2\pi}d\sigma^1 \tanh^2\rho =\frac{\sqrt{\mathfrak{g}}}{8\pi}(\Lambda_{1}(\rho_m)+\omega \rho_m \Lambda_{2}(\rho_m)+\cdots)
\label{e26}
\end{eqnarray}
 where we identify each of the individual entities as\footnote{Here $ B_{z}(a,b) $ is the \emph{incomplete} beta function which becomes the usual beta function for $ z=1 $ \cite{abramh}.},
 \begin{eqnarray}
 \Lambda_{1}(\rho_m)=\left[ B_{-e^{2 \rho_m \sinh ^{-1}(2 \pi )}}\left(-\frac{1}{2 \rho_m},0\right)+B_{-e^{2 \rho_m \sinh ^{-1}(2 \pi )}}\left(1-\frac{1}{2 \rho_m},0\right)\right] \nonumber\\
 \times \left(\sqrt{1+4 \pi ^2}-2 \pi \right) \left(-e^{2\rho_m \sinh ^{-1}(2 \pi )}\right)^{\frac{1}{2 \rho_m}}\frac{1}{4 \rho_m^2}\nonumber\\
 -\left[B_{-e^{2\rho_m \sinh ^{-1}(2 \pi )}}\left(\frac{1}{2 \rho_m},0\right)+B_{-e^{2 \rho_m \sinh ^{-1}(2 \pi )}}\left(1+\frac{1}{2 \rho_m},0\right) \right] \nonumber\\
\times  \left(\sqrt{1+4 \pi ^2}+2 \pi \right) \left(-e^{2  \rho_m \sinh ^{-1}(2 \pi )}\right)^{-\frac{1}{2  \rho_m}}\frac{1}{4 \rho_m^2}\nonumber\\
+\frac{1}{4 \rho_m^2}\left( 8 \pi \rho_m^2 +\pi  \tan \left(\frac{\pi }{4 \rho_m}\right)+\pi  \cot \left(\frac{\pi }{4 \rho_m}\right)\right) \nonumber\\
-\frac{ \sqrt{1+4 \pi ^2}}{\rho_m}  \tanh \left(\rho_m \sinh ^{-1}(2 \pi )\right)
\label{E29}
 \end{eqnarray}
and,
\begin{eqnarray}
 \Lambda_{2}(\rho_m)=\int_0^{2\pi}d\sigma^1 \frac{\tanh \left(\rho_m \sinh ^{-1}\sigma^1 \right)}{\sqrt{(\sigma^1) ^2+1}}~~~~~~~~~~~~~~~~~~~~~~~~~~~~~~~~~~~~~~~~~~~~~~\nonumber\\
 \times \left( \tan ^{-1}\sigma^1 -2 \sigma^1  \tanh ^2\left(\rho_m \sinh ^{-1}\sigma^1 \right)- \tan ^{-1}\sigma^1 \tanh ^2\left(\rho_m \sinh ^{-1}\sigma^1\right) \right) +\mathcal{O}(1/\rho_m).
\end{eqnarray}

In the large $ \rho_m (\gg1) $ limit, the above charges (\ref{e23}) and (\ref{e26}) become,
\begin{eqnarray}
\label{e29}
E_{NR} & \approx & \frac{\sqrt{\mathfrak{g}}}{8\pi}\rho_m^2 (\tilde{\Delta}_1 +\omega \rho_m  \tilde{\Delta}_{2}+\cdots) \\
S_{\varphi}& \approx & \frac{\sqrt{\mathfrak{g}}}{8\pi} \omega \rho_m \tilde{\Lambda}_2 
\label{e30}
\end{eqnarray}
where we denote, $ \Delta_1\Big|_{\rho_m \gg 1} = \rho^2_m \tilde{\Delta}_1  $ and so on. 

Combining (\ref{e29}) and (\ref{e30}) we finally obtain,
\begin{eqnarray}
E_{NR}  & \sim &S_{\varphi}+(8\pi \tilde{\Lambda}_2^{-2}) ~\sqrt{\mathfrak{g}} \left[  \tilde{\Delta}_1  \left( \frac{S_{\varphi}}{\sqrt{\mathfrak{g}}}\right)^2 + (8 \pi \tilde{\Delta}_{2}\tilde{\Lambda}_2^{-1})\left( \frac{S_{\varphi}}{\sqrt{\mathfrak{g}}}\right)^3+\cdots\right] 
\end{eqnarray}
where we identify, $\frac{S_{\varphi}}{\sqrt{\mathfrak{g}}}(< 1 )$ as the effective spin in the strong coupling regime of SMT.
\subsection{Spinning strings in $S^5 $}
\subsubsection{Single spin solution}
The second embedding that we consider is that of a NR folded spinning string on $ S^5 $ whose centre of mass is fixed and located at the north pole. The string soliton is considered to be stretched along one of the polar coordinates of $ S^5 $ whose end points are spinning along the azimuthal direction $ \phi_1 $,
\begin{eqnarray}
t = \sigma^0 ~;~\eta = \sigma^1 ~;~\rho = 0~;~ \theta_1 = \theta_1 (\sigma^1)~;~\phi_1 = \omega \sigma^0
\end{eqnarray}
while all the remaining coordinates are switched off.

The corresponding sigma model Lagrangian turns out to be,
\begin{eqnarray}
\mathcal{L}_{NG}=\frac{\omega}{2}\sin^2\theta_1 +\frac{\theta'^2_1}{2}.
\end{eqnarray}

The resulting equation of motion,
\begin{eqnarray}
\theta''_1 -\frac{\omega}{2}\sin2\theta_1 =0
\end{eqnarray}
could be integrated once to obtain,
\begin{eqnarray}
\theta'^2_1 =\omega (\cos^2\theta_m -\cos^2\theta_1).
\end{eqnarray}

The space time energy of the NR string is given by\footnote{Here, $ F(\varphi | k^2) $ and $ E (\varphi | k^2)$ are respectively the incomplete \emph{elliptic integrals} of the first and second kind \cite{abramh}.},
\begin{eqnarray}
E_{NR} &=& \frac{\sqrt{\mathfrak{g}}}{ 2\pi}\sqrt{\omega}\int_0^{\theta_m}d \theta_1~ \sqrt{(\cos^2\theta_m -\cos^2\theta_1)} \nonumber\\
&=&\frac{\sqrt{\mathfrak{g}}}{ 2\pi}\sqrt{\omega}\sin\theta_m ~E\left(\theta_m \left|\csc ^2\theta_m \right.\right).
\label{e35} 
\end{eqnarray}

On the other hand, the R- charge on the stringy side can be computed as,
\begin{eqnarray}
J&=&\frac{\sqrt{\mathfrak{g}}}{2 \pi}\int_0^{\theta_m}\frac{d \theta_1}{\theta'_1}~ \sin^2\theta_1 \nonumber\\
& = & \frac{\sqrt{\mathfrak{g}}}{2 \pi \sqrt{\omega}}\sin\theta_m (E\left(\theta_m \left|\csc ^2\theta_m\right.\right)-F\left(\theta_m \left|\csc ^2\theta_m\right.\right)).
\label{e36}
\end{eqnarray}

Combining both (\ref{e35}) and (\ref{e36}) we find,
\begin{eqnarray}
E_{NR} =\omega J +\frac{\sqrt{\mathfrak{g}}}{2 \pi}\sqrt{\omega}\sin\theta_m F\left(\theta_m \left|\csc ^2\theta_m\right.\right).
\end{eqnarray}

As a further remark, we notice that,
\begin{eqnarray}
2 \pi =4 \int_0^{\theta_m}\frac{d \theta_1}{\theta'_1}=\frac{4}{\sqrt{\omega}}\frac{1}{\sin\theta_m}F\left(\theta_m \left|\csc ^2\theta_m\right.\right).
\label{e39}
\end{eqnarray}

Using (\ref{e39}) we finally obtain\footnote{We set the overall scale factor, $ \omega =1 $.},
\begin{eqnarray}
E_{NR} \sim J +\frac{\sqrt{\mathfrak{g}}}{4}\sin^2\theta_m
\end{eqnarray}
which clearly reveals that the maximum correction to the anomalous dimension ($ \Delta_{NR} $) in the dual SMT is $ \sim \frac{\sqrt{\mathfrak{g}}}{4} $ which corresponds to spinning strings whose end points are stretched upto the equatorial plane ($ \theta_m = \frac{\pi}{2} $) of $ S^5 $. 
\subsubsection{Multi-spin solutions} 
We now generalize the above result for NR folded (closed) spinning string configurations with two unequal spins (to start with) along two azimuthal directions \cite{Frolov:2003xy},
\begin{eqnarray}
t = \sigma^0 ~;~\eta = \sigma^1 ~;~\rho = 0~;~\theta_2 = 0~;~ \theta_1 = \theta_1 (\sigma^1)~;~ \phi_1 = \omega_{1} \sigma^0 ~;~\phi_2 = \omega_{2} \sigma^0.
\end{eqnarray}

The corresponding Lagrangian density turns out to be,
\begin{eqnarray}
\mathcal{L}_{NG}=\frac{\omega_1}{2}\sin^2\theta_1 - \omega_2 \left(\frac{1}{3}-\frac{1}{2}\sin^2\theta_1 \right) +\frac{\theta'^2_1}{2}
\end{eqnarray}
which yields the equation of motion of the form,
\begin{eqnarray}
\theta''_1 -\frac{\tilde{\omega}}{2}\sin2\theta_1 =0
\end{eqnarray}
where, $ \tilde{\omega}=\omega_1 +\omega_2 $ is the total angular frequency of the string.

The corresponding conserved charges are given by,
\begin{eqnarray}
E_{NR} &=&\frac{\sqrt{\mathfrak{g}}}{ 2\pi}\sqrt{\tilde{\omega}}\sin\theta_m ~E\left(\theta_m \left|\csc ^2\theta_m \right.\right)\\
J_1&=&\frac{\sqrt{\mathfrak{g}}\sin\theta_m}{2 \pi \sqrt{\tilde{\omega}}} (E\left(\theta_m \left|\csc ^2\theta_m\right.\right)-F\left(\theta_m \left|\csc ^2\theta_m\right.\right))\\
J_2 &=&\frac{\sqrt{\mathfrak{g}}\sin\theta_m}{6\pi \sqrt{\tilde{\omega}}}\left( \left(3-2 \csc ^2\theta_m \right) F\left(\theta_m\left|\csc ^2\theta_m\right.\right)-3 E\left(\theta_m\left|\csc ^2\theta_m \right.\right)\right).
\end{eqnarray}

A straightforward computation reveals the total R- charge of the configuration as,
\begin{eqnarray}
|J|=|J_1 +J_2| =\frac{\sqrt{\mathfrak{g}}}{3\pi \sqrt{\tilde{\omega}}}(\sin\theta_m)^{-1}F\left(\theta_m\left|\csc ^2\theta_m\right.\right)=\frac{\sqrt{\mathfrak{g}}}{6}.
\end{eqnarray}

Combining all these results, we finally obtain,
\begin{eqnarray}
E_{NR}=J+\frac{\sqrt{\mathfrak{g}}}{2 \pi}\left(\sqrt{\tilde{\omega}}\sin\theta_m ~E\left(\theta_m \left|\csc ^2\theta_m \right.\right)+\frac{\pi}{3} \right). 
\end{eqnarray}

Setting, $ \omega_1=\omega_2=1 $ and $ \theta_m = \frac{\pi}{2} $ we find the leading order correction to the spectrum,
\begin{eqnarray}
E_{NR}=J +\gamma ~\sqrt{\mathfrak{g}}~;~\gamma =0.391746
\end{eqnarray}
Notice that the value of the correction coefficient ($ \gamma $) increases from single spin solution ($ \gamma = 0.25 $) to multi spin solution by 14\%.
\subsection{Folded three spin solutions}
We now generalize the above construction by considering NR folded spinning strings in $AdS_5\times S^5 $ with two equal spins in $ AdS_5 $ and one spin along $ S^5 $,
\begin{eqnarray}
t = \sigma^0 ~;~\eta = \sigma^1~;~\rho  = \rho (\sigma^1)~;~ \psi = 0~;~\theta_1 = \theta_1 (\sigma^1)~;~\chi = \varphi =\omega \sigma^0 ~;~\phi_1 = \nu \sigma^0
\end{eqnarray}
where the string is stretched along the radial coordinate $ \rho $ of $ AdS_5 $ as well as the angular direction $ \theta_1 $ of $ S^5 $.

The corresponding Lagrangian density turns out to be,
\begin{eqnarray}
\mathcal{L}_{NG}=-\frac{3 \omega}{2}\tanh^2\rho +\frac{\nu}{2}\cosh^{-2}\rho\sin^2\theta_1 +\frac{1}{2}\cosh^2\rho (\rho'^2 + \theta'^2_1).
\end{eqnarray}

The resulting equations of motion are given by,
\begin{eqnarray}
\rho'' \cosh^2 \rho +(\rho'^2 - \theta'^2_1)\cosh\rho \sinh\rho \nonumber\\
+ 3\omega \tanh\rho \cosh^{-2}\rho + \nu \cosh^{-3}\rho \sinh\rho \sin^2\theta_1 =0
\label{e54}
\end{eqnarray}
and,
\begin{eqnarray}
\theta_1'' \cosh^2\rho + \theta'_1 \rho' \sinh2\rho - \frac{\nu}{2} \cosh^{-2}\rho \sin2\theta_1 =0.
\label{e55}
\end{eqnarray}
\subsubsection{Perturbative solution}
The above set of equations (\ref{e54})-(\ref{e55}) are indeed difficult to solve analytically. To simplify the problem, we consider a configuration where the string soliton is considered to be sitting at the equatorial plane ($ \theta_1 = \frac{\pi}{2} $) of $ S^5 $ while having a spin along the azimuthal direction $ \phi_1 $. This clearly satisfies (\ref{e55}) and upon substitution into (\ref{e54}) yields,
\begin{eqnarray}
\rho'' \cosh^2 \rho +\rho'^2 \cosh\rho \sinh\rho 
+ 3\omega \tanh\rho \cosh^{-2}\rho + \nu \cosh^{-3}\rho \sinh\rho =0.
\label{e56}
\end{eqnarray}

We propose the following perturbative solutions for\footnote{Like before, we take into account the \emph{slow frequency} limit for strings which amounts of considering, $ \omega \rho_m \ll 1 $ and $ \nu \rho_m \ll 1 $.} (\ref{e56}),
\begin{eqnarray}
\rho &=& \rho^{(0)}(1+\omega \rho_{(\omega)}+\nu \rho_{(\nu)}+\cdots )
\end{eqnarray}
where the zeroth order solution we note down to be,
\begin{eqnarray}
\rho^{(0)}(\sigma^1)=\rho_m \sinh ^{-1}\sigma^1.
\end{eqnarray}

The leading order equation in $ \omega $ has a solution of the form,
\begin{eqnarray}
 \rho_{(\omega)}(\sigma^1)=\frac{2 \sigma^1 +3 \tan ^{-1}\sigma^1}{2 \sqrt{(\sigma^1) ^2+1} \sinh ^{-1}\sigma^1}.
\end{eqnarray}

Finally, the leading order solution in $ \nu $ turns out to be,
\begin{eqnarray}
\rho_{(\nu)}(\sigma^1)=\frac{2 \sigma^1 +\tan ^{-1}\sigma^1}{2 \sqrt{(\sigma^1) ^2+1} \sinh ^{-1}\sigma^1}.
\end{eqnarray}
\subsubsection{Energy-spin relations}
Energy of the stringy configuration turns out to be,
\begin{eqnarray}
E_{NR}&=&\frac{\sqrt{\mathfrak{g}}}{8\pi}\int_{0}^{2\pi}d\sigma^1 \rho'^2 \cosh^2\rho\nonumber\\
& \simeq & \frac{\sqrt{\mathfrak{g}}}{8\pi}(\Delta_1 (\rho_m)+\omega \rho_m^2 \Sigma_{(\omega)}(\rho_m)+\nu \rho^2_m \Sigma_{(\nu)} (\rho_m)+\cdots)
\label{e61}
\end{eqnarray}
where the sub-leading corrections in (\ref{e61}) are given by,
\begin{eqnarray}
 \Sigma_{(\nu)} (\rho_m)=\int_0^{2 \pi}\frac{d\sigma^1 \cosh \left(\rho_m \sinh ^{-1}\sigma^1\right)}{\left((\sigma^1) ^2+1\right)^{5/2}}\times \mathcal{I}_1(\sigma^1 , \rho_m)\\
 \mathcal{I}(\sigma^1 , \rho_m)=\sqrt{(\sigma^1) ^2+1} \left(3-\sigma^1  \tan ^{-1}\sigma^1\right) \cosh \left(\rho_m \sinh ^{-1}\sigma^1\right)\nonumber\\
 +\rho_m \left((\sigma^1) ^2+1\right) \left(2 \sigma^1 +\tan ^{-1}\sigma^1 \right) \sinh \left(\rho_m \sinh ^{-1}\sigma^1 \right)
\end{eqnarray}
and,
\begin{eqnarray}
\Sigma_{(\omega)}(\rho_m)=\int_0^{2 \pi}d \sigma^1\cosh \left(\rho_m \sinh ^{-1}\sigma^1\right)\times \mathcal{I}_2 (\sigma^1 , \rho_m)\\
\mathcal{I}_2(\sigma^1 , \rho_m)=\frac{\left(5-3 \sigma^1  \tan ^{-1}\sigma^1 \right) \cosh \left(\rho_m \sinh ^{-1}\sigma^1 \right)}{\left((\sigma^1)^2+1\right)^2}\nonumber\\
+\frac{\rho_m \left(2 \sigma^1 +3 \tan ^{-1}\sigma^1 \right) \sinh \left(\rho_m \sinh ^{-1}\sigma^1 \right)}{\left((\sigma^1) ^2+1\right)^{3/2}}.
\end{eqnarray}

The total spin of the configuration, on the other hand, is given by
\begin{eqnarray}
S &=& -\frac{3\sqrt{\mathfrak{g}}}{8\pi}\int_{0}^{2\pi}d\sigma^1 \tanh^2 \rho \nonumber\\
&\simeq & -\frac{3\sqrt{\mathfrak{g}}}{8\pi}(\Lambda_{1}(\rho_m)+\omega \rho_m \Psi_{(\omega)}(\rho_m)+\nu \rho_m \Psi_{(\nu)}(\rho_m)+\cdots)
\end{eqnarray}
where, the sub-leading corrections are given by,
\begin{eqnarray}
\Psi_{(\omega)}(\rho_m)&=&\int_0^{2 \pi } \frac{ \left(2 \sigma^1 +3 \tan ^{-1}\sigma^1\right) \tanh \left(\rho_m \sinh ^{-1}\sigma^1 \right) \text{sech}^2\left(\rho_m \sinh ^{-1}\sigma^1 \right)}{\sqrt{(\sigma^1) ^2+1}} \, d\sigma^1\\
\Psi_{(\nu)} (\rho_m)&=&\int_0^{2 \pi } \frac{ \left(2 \sigma^1 + \tan ^{-1}\sigma^1\right) \tanh \left(\rho_m \sinh ^{-1}\sigma^1 \right) \text{sech}^2\left(\rho_m \sinh ^{-1}\sigma^1 \right)}{\sqrt{(\sigma^1) ^2+1}} \, d\sigma^1.
\end{eqnarray}

Finally, the R-charge of the configuration is given by
\begin{eqnarray}
J&=&  \frac{\sqrt{\mathfrak{g}}}{8\pi}\int_{0}^{2\pi}d\sigma^1 \cosh^{-2}\rho\nonumber\\
& \simeq & \frac{\sqrt{\mathfrak{g}}}{8\pi} (\Phi(\rho_m)-\omega \rho_m \Psi_{(\omega)}(\rho_m)-\nu \rho_m \Psi_{(\nu)}(\rho_m)+\cdots)
\end{eqnarray}
where, $ \Phi(\rho_m) $ is the leading order correction to the R-charge that goes like $ \sim \frac{1}{\rho^2_m} $ similar to that of $ \Lambda_{1}(\rho_m) $ as given in (\ref{E29}).

Considering the extended string ($ \rho_m \gg 1$) limit, the total angular momentum of the NR stringy configuration turns out to be,
\begin{eqnarray}
Q =  |S+J| \approx \frac{\sqrt{\mathfrak{g}}}{2 \pi}(\omega \rho_m \Psi_{(\omega)}(\rho_m)+\nu \rho_m \Psi_{(\nu)}(\rho_m)+\cdots)
\end{eqnarray}
which finally results in the dispersion relation of the form,
\begin{eqnarray}
E_{NR} \sim Q +\frac{\sqrt{\mathfrak{g}}\rho^2_m}{8 \pi}(\tilde{\Delta}_1 (\rho_m)+\omega  \Sigma_{(\omega)}(\rho_m)+\nu  \Sigma_{(\nu)} (\rho_m)+\cdots) +\mathcal{O}(\sqrt{\mathfrak{g}}/\rho_m).
\end{eqnarray}
\section{Spinning strings in $ R \times S^3 $}
The purpose of this Section is to revisit the NR spinning string configurations and establish a precise mapping between NR sigma models in $ AdS_5 \times S^5 $ and one dimensional Neumann-Rosochatius like integrable models \cite{Babelon:1992rb}. We select a specific sub-sector $ R \times S^3\subset AdS_5 \times S^5 $ of the full target space geometry where we consider the string soliton to be sitting at the centre of $ AdS_5 $ and switch on spin along one of the azimuthal directions ($ \phi_2 $) of $ S^5 $. 

We choose the string embedding of the following form,
\begin{eqnarray}
t = \sigma^0 ~;~\eta = \sigma^1 ~;~\rho =0~;~\theta_2 =\pi/2 ~;~ \theta_1 = \theta_1 (\sigma^1 )~;~\phi_i = \phi_i (\sigma^{\alpha})
\end{eqnarray}
and switch off all the remaining coordinates. For simplicity, from now on we set, $ \theta_1 = \theta $.

The resulting sigma model Lagrangian turns out to be,
\begin{eqnarray}
\mathcal{L}_{NG}=\left(\frac{1}{2}\sin^2\theta - \frac{1}{3} \right) \dot{\phi}_2 +\frac{\theta'^2}{2}-\frac{1}{8}\cos^4\theta \phi'^2_2 \nonumber\\
+\frac{1}{8}\left(\sin^2\theta \phi'^2_1 + \cos^2\theta \phi'^2_2 \right).
\label{e73}
\end{eqnarray}

To proceed further, we redefine coordinates as,
\begin{eqnarray}
\ell_1 =\frac{1}{\sqrt{2}} \sin\theta ~;~ \ell_2 = \frac{1}{\sqrt{2}}\cos\theta  ~;~ \phi_1 = \sqrt{8}\xi_{1}(\sigma^1)~;~ \phi_2 = - \nu \sigma^0 +\sqrt{8}\xi_{2}(\sigma^1)
\end{eqnarray}
which upon substitution into (\ref{e73}) yields,
\begin{eqnarray}
\mathcal{L}_{1D}=\ell'^2_i + \ell^2_i \xi'^2_i - \ell^2_i \nu \delta_{i1}- \mathcal{N}(\ell^2_i -1/2)+\Delta_{S^5} .
\label{e75}
\end{eqnarray}

The first four terms on the R.H.S of (\ref{e75}) together constitute what we define as the nonrelativistic analogue of 1D Neumann-Rosochatius like integrable models derived directly from 2D sigma models in the $ SU(1,2|3) $ SMT limit of strings on $ AdS_5 \times S^5 $. Contrary to its relativistic cousins \cite{Arutyunov:2003uj}-\cite{Arutyunov:2003za}, the nonrelativistic 1D model (\ref{e75}) is linear in the time derivative/ frequency ($ \nu $) and is quadratic in space derivatives. Here, $ \mathcal{N} $ is the Lagrange multiplier that preserves the constraint condition, $ \ell^2_i =1 $. 

The last term on the R.H.S. of (\ref{e75}),
\begin{eqnarray}
\Delta_{S^5} =\frac{\nu}{3}-4 \ell^4_i \xi'^2_i \delta_{i2}
\end{eqnarray}
is identified as the deformation to the nonrelativistic Neumann-Rosochatius model.

The phase space of the 1D model (\ref{e75}) is eventually 6 ($ =2N $) dimensional. The configuration is integrable as there are $ \mathcal{Q}_i~(i=1,2) $ conserved charges associated with the dynamical phase space. This follows directly from the fact that the system is constrained.

The canonical momentum densities are given by,
\begin{eqnarray}
\Pi_{\ell_i}&=&2 \ell'_i \\
\Pi_{\xi_1}& = & 2 \ell^2_1 \xi'_1 = \upsilon_{1} \\
\Pi_{\xi_2}& = & 2 \ell^2_2 \xi'_2 (1 - 4 \ell^2_2)= \upsilon_{2}.
\end{eqnarray}

Clearly, there are two constants of motion ($ \upsilon_{i} \sim \mathcal{Q}_i$) associated with the configuration. The corresponding canonical Hamiltonian density turns out to be,
\begin{eqnarray}
\mathcal{H}_{1D} = \nu \mathcal{I}_1
\end{eqnarray}
where we identify,
\begin{eqnarray}
\mathcal{I}_1 = \ell^2_1 -\frac{1}{3} +\frac{1}{4 \nu}\left(\frac{\upsilon_{1}^2}{\ell^2_1}+\frac{\upsilon^2_2}{\ell^2_2 (1- 4\ell^2_2)} +(\ell_1 \ell'_2 -\ell_2 \ell'_1)^2\right) 
\end{eqnarray}
as nonrelativistic analogue of the Uhlenbeck constant \cite{Arutyunov:2003uj}-\cite{Arutyunov:2003za}.
\subsection{Constant radii solutions}
We first note down the equations of motion that directly result from (\ref{e75}),
\begin{eqnarray}
\label{e83}
\ell''_1 - \ell_1 \xi'^2_1 + \ell_1 (\nu + \mathcal{N}) &=&0\\
\ell''_2 - \ell_2 \xi'^2_2 (1- 8 \ell^2_2) +\mathcal{N} \ell_2  &=&0.
\label{e84}
\end{eqnarray}

The above set of equations (\ref{e83})-(\ref{e84}) could in principle be solved considering a fixed radius \cite{Arutyunov:2003uj} ansatz namely, $ \ell_i = \mathfrak{a}_i =$ constant and choosing an ansatz for the winding modes along the azimuthal direction of $ S^5 $,
\begin{eqnarray}
\xi_i (\sigma^1)= \mathfrak{m}_i ~ \sigma^1
\end{eqnarray}
where $ \mathfrak{m}_i $s are the respective winding numbers. 

This results in the following set of algebraic equations,
\begin{eqnarray}
\label{e86}
\mathfrak{m}^2_1 - \nu - \mathcal{N}& =& 0\\
\mathfrak{m}^2_2 (1- 8 \mathfrak{a}^2_2)- \mathcal{N}&=&0.
\label{e87}
\end{eqnarray}

Using (\ref{e86})-(\ref{e87}), we finally obtain,
\begin{eqnarray}
\mathfrak{a}^2_1 &=& \frac{7 \mathfrak{m}^2_2 + \mathfrak{m}^2_1 -\nu}{8 \mathfrak{m}^2_2}\\
\mathfrak{a}^2_2 &=& \frac{ \mathfrak{m}^2_2 - \mathfrak{m}^2_1 +\nu}{8 \mathfrak{m}^2_2}
\end{eqnarray}
subjected to the constraint condition, $ \mathfrak{a}^2_1 + \mathfrak{a}^2_2 =1 $.
\subsection{Dispersion relation}
The energy of the NR stringy configuration is given by,
\begin{eqnarray}
E_{NR}=\frac{\sqrt{\mathfrak{g}}}{4}\mathfrak{a}^2_1 (\mathfrak{m}^2_1 +\mathfrak{a}^2_2 \mathfrak{m}^2_2).
\label{e90}
\end{eqnarray}

On the other hand, the R- charge of the configuration reads as,
\begin{eqnarray}
J = \frac{\sqrt{\mathfrak{g}}}{2}(\mathfrak{a}^2_1 -\frac{1}{3}).
\label{e91}
\end{eqnarray}

Combining (\ref{e90}) and (\ref{e91}) and after some trivial algebra we find,
\begin{eqnarray}
E_{NR}\sim \left( \frac{\mathfrak{m}^2_1}{2}+\frac{\mathfrak{m}^2_2}{3}\right) J +\frac{\sqrt{\mathfrak{g}}}{2}\left(\mathfrak{m}^2_1 +\frac{2}{3}\mathfrak{m}^2_2 - 2\mathfrak{m}^2_2 (\tilde{J}+6 \tilde{J}^2) \right) 
\end{eqnarray}
where, we define $ \tilde{J}=\frac{J}{\sqrt{\mathfrak{g}}} $ as the effective R- charge in strong coupling limit of SMT. 
\section{Spinning strings in $ AdS_5 $}
We now focus on \emph{folded} spinning strings in $ AdS_5 $ and switch off any dynamics along $ S^5 $. In order to map the corresponding sigma model action into a 1D Neumann-Rosochatius like integrable model we choose to work with the following string embedding,
\begin{eqnarray}
t = \sigma^0 ~;~ \eta = \sigma^1 ~;~ \psi \sim 0~;~ \rho = \rho (\sigma^1)~;~ \chi = const.~;~ \varphi = \varphi (\sigma^{\alpha}).
\end{eqnarray}

The resulting Lagrangian density turns out to be,
\begin{eqnarray}
\mathcal{L}_{NG} \simeq -\tanh^2 \rho \dot{\varphi}+\cosh^2 \rho \rho'^2 +\frac{1}{4}\sinh^2 \rho \varphi'^2.
\end{eqnarray}

We define the following set of coordinates,
\begin{eqnarray}
z_0 = \cosh\rho ~;~z_1 = \sinh\rho ~;~\varphi = \omega \sigma^0 +2 \beta (\sigma^1)
\label{e96}
\end{eqnarray}
and consider the dynamics near the centre of $ AdS_5 $. 

This finally results in the 1D Lagrangian of the following form,
\begin{eqnarray}
\mathcal{L}_{1D}\sim g^{ab}(z'_a z'_b + z_a z_b \beta'^2 -\omega z_a z_b)+ \mathcal{G} (g^{ab}z_a z_b + 1) +\Delta_{AdS}
\label{e97}
\end{eqnarray}
where, we introduce the diagonal metric $ g^{ab}= diag (-1 , 1) $ as well as the Lagrange multiplier $ \mathcal{G} $ which implies the constraint condition \cite{Arutyunov:2003za},
\begin{eqnarray}
g^{ab}z_a z_b = -1.
\label{e98}
\end{eqnarray}

The first two terms (in the parenthesis) on the R.H.S. of  (\ref{e97}) are the standard Neumann-Rosochatius piece while the remaining one,
\begin{eqnarray}
\Delta_{AdS} = - \omega + \beta'^2
\end{eqnarray}
serves as an extension to it.
\subsection{Equations of motion}
The equations of motion could be enumerated as,
\begin{eqnarray}
z''_a - z_a \beta'^2 +(\omega - \mathcal{G})z_a =0~;~a=0,1.
\end{eqnarray}

To obtain solutions, we first set, $ \beta (\sigma^1)= k \sigma^1 $ where $ k $ is the winding number of the string along $ \varphi $. The corresponding solutions turn out to be,
\begin{eqnarray}
z_0 =\cosh\sqrt{p}\sigma^1 ~;~ z_1 =\sinh\sqrt{p}\sigma^1 ~;~p=k^2 - \omega >0 
\label{e101}
\end{eqnarray}
which satisfy the constraint (\ref{e98}).

A direct comparison between (\ref{e96}) and (\ref{e101}) further reveals,
\begin{eqnarray}
\rho (\sigma^1) = \sqrt{p} ~\sigma^1.
\end{eqnarray}

\subsection{Energy-spin relation}
The energy and as well as the spin of the NR configuration are given below,
\begin{eqnarray}
E_{NR}&=&\frac{\sqrt{\mathfrak{g}}}{\pi \sqrt{p}}\int_0^{\rho_m}d \rho (p \cosh^2\rho +k^2 \sinh^2\rho)\nonumber\\
&=&\frac{\sqrt{\mathfrak{g}}}{4\pi \sqrt{k^2 - \omega}}\left(-2 \omega \rho_m +\left(2k^2 - \omega \right) \sinh (2 \rho_m)\right)\nonumber\\
& \approx & \frac{\sqrt{\mathfrak{g}}}{\pi \sqrt{k^2 - \omega}}\left((k^2 - \omega)\rho_m + \frac{\rho_m^3}{3}  \left(2k^2 - \omega \right)\right),
\label{e103}
\end{eqnarray}
\begin{eqnarray}
S_{\varphi} &=&\frac{\sqrt{\mathfrak{g}}}{\pi \sqrt{p}}\int_0^{\rho_m}d \rho \tanh^2\rho \nonumber\\
&=&\frac{\sqrt{\mathfrak{g}}}{\pi \sqrt{k^2 - \omega}}(\rho_m -\tanh\rho_m)\nonumber\\
& \approx & \frac{\sqrt{\mathfrak{g}}\rho^3_m}{3\pi \sqrt{k^2 - \omega}}.
\label{e104}
\end{eqnarray}

Using (\ref{e103}) and (\ref{e104}) we finally obtain,
\begin{eqnarray}
E_{NR} \sim (2 k^2 - \omega)S_{\varphi} +\sqrt{\mathfrak{g}}\left( \frac{3}{\pi^2}(k^2 - \omega)^2\right)^{1/3} \left( \frac{S_{\varphi}}{\sqrt{\mathfrak{g}}}\right)^{1/3} +\cdots
\end{eqnarray}
where, $  \frac{S_{\varphi}}{\sqrt{\mathfrak{g}}} \sim \rho^3_m \ll 1 $ is the effective spin in the strong coupling regime of SMT.
\section{Summary and final remarks}
We conclude our paper with a brief summary of the analysis and mentioning some of its possible implications as well. The present paper explores various corners of the $ SU(1,2|3) $ Spin-Matrix theory (SMT) limit of $ \mathcal{N}=4 $ SYM in the limit of strong coupling. Using the dual \emph{semiclassical} nonrelativistic stringy counterpart in $ AdS_5 \times S^5 $, the present paper realizes the spectrum of the theory for various sub-sectors of the full Hilbert space. We also show that the $ SU(1,2|3) $ Spin-Matrix theory (SMT) limit of the sigma model in $ AdS_5 \times S^5 $ can be mapped into 1D Neumann-Rosochatius like integrable models. This further allows us to compute the spectrum of the 2D sigma model corresponding to various spinning string configurations in $ AdS_5 \times S^5 $. 

It would be really nice to reproduce these semiclassical string states (and hence the spectrum) considering a large $ \mathfrak{g}(\gg1 $) Spin-Matrix theory limit in the $ SU(1,2|3) $  sector of $ \mathcal{N}=4 $ SYM. Below we outline some possible steps in order to achieve this goal in the limit $ N\rightarrow \infty $. The nonrelativistic spinning strings considered in this paper probe various subsectors of SMT (with largest possible Hilbert space) that can be obtained as a decoupling limit \cite{Harmark:2006di}, \cite{Harmark:2007px}, \cite{Harmark:2014mpa} of $ \mathcal{N}=4 $ SYM on $ \mathbb{R}\times S^3 $. For example, nonrelativistic spinning string states in $ AdS_5 $ correspond to the decoupling limit in the sector spanned by the single trace operators of the form $ \mathcal{O}_{S}\sim tr (\Phi_Z (n^{\mu}D_{\mu})^S \Phi_Z) $ in $ \mathcal{N}=4 $ SYM. Here, $ \Phi_Z $s are the complex scalars built out of two of the six real scalars in $ \mathcal{N}=4 $ SYM, $ D_{\mu} $ is the gauge covariant derivative and $ n^{\mu} $ is the constant null vector \cite{Frolov:2003qc}. On the other hand, nonrelativistic multispin string states on $ S^5 $ can be realised as the decoupling limit of the sector (in $ \mathcal{N}=4 $ SYM) spanned by the gauge invariant (single trace) operators of the form, $ \mathcal{O}_{J_1,J_2}\sim tr (\Phi_Z^{J_1}\Phi_X^{J_2}) +$ (permutations). Here, for example, the decoupling/quantum mechanical limit corresponds to setting $ (T,\omega_1,\omega_2, \Omega_1 ,\Omega_2,\Omega_3)=(0,0,0,1,1,0) $ while keeping the ratios $ \tilde{T}=\frac{T}{1-\Omega} $ and $ \tilde{\lambda}=\frac{\lambda}{1-\Omega} $ fixed  \cite{Harmark:2006di}. Here, $ \Omega_1 $ and $ \Omega_2 $ are the chemical potentials corresponding to the R-charges $ J_1 $ and $ J_2 $. The resulting states belong to $ SU(2) \subset SU(1,2|3)$ representation of the underlying spin group in SMT \cite{Harmark:2014mpa}. Finally, nonrelativistic spinning string states with two spins along $ S^3\subset AdS_5 $ and one spin along $ S^5 $ can be realised as a quantum mechanical limit of the sector spanned by the gauge invariant operators of the form, $ \mathcal{O}_{S_1 , S_2, J} \sim tr (\Phi_Z (n^{\mu}D_{\mu})^{S_{1}} (n^{\nu}D_{\nu})^{S_{2}}\Phi_Z \Phi_{X}^{J-2})+\cdots$ with classical dimension $ \Delta_0 \sim Q \sim S_1 + S_2 +J $. The quantum mechanical limit for these operators corresponds to setting $ (T,\omega_1,\omega_2, \Omega_1, \Omega_2, \Omega_3)=(0,1,1,1,0,0) $ which represents states in the $ SU(1,2|2)\subset SU(1,2|3) $ representation of the underlying spin group \cite{Harmark:2014mpa}. 

The one loop correction to the respective dilatation operators would essentially serve as the interaction Hamiltonian ($ H_{int} $) \cite{Harmark:2014mpa} for the corresponding quantum mechanical theory in the near critical region. Considering the planar ($ N\rightarrow \infty $) limit, the final step would be to map this quantum mechanical Hamiltonian ($ H_{int} $) into a one dimensional periodic spin-chain \cite{Harmark:2006di}, \cite{Harmark:2008gm}, \cite{Harmark:2014mpa}. In the regime of strong ($ \mathfrak{g}\gg 1 $) coupling and considering large values of the corresponding Cartan/R- symmetry generators ($ Q$), the spectrum associated with this interaction Hamiltonian ($ H_{int} $) may be obtained as an expansion in the ratio involving both the SMT coupling $ (\mathfrak{g})$ and $ Q $ \cite{Harmark:2014mpa} and thereby can be compared with nonrelativistic semiclassical string spectrum obtained in this paper. Going back to the periodic spin chain description (with added \emph{impurities}), it may also be worthwhile in trying to diagonalize the interaction Hamiltonian ($ H_{int} $) using the powerful techniques of integrability \cite{Beisert:2003xu} and thereby comparing the spectrum on both sides of the duality. We hope to carry out a detailed analysis on some of these issues in the near future.\\  

{\bf {Acknowledgements :}}
 The author is indebted to the authorities of IIT Roorkee for their unconditional support towards researches in basic sciences. 

\end{document}